\documentclass[twocolumn,showpacs,preprintnumbers,amsmath,amssymb]{revtex4}

\usepackage{graphicx}
\usepackage{amsfonts}
\usepackage{amsmath}

\newcommand{\EP}{{\scriptscriptstyle\mathrm{EP}}}

\begin{document}

\title{Geometric phase around exceptional points}

\author{Alexei A. Mailybaev}
\email{mailybaev@imec.msu.ru}
\author{Oleg N. Kirillov}
\email{kirillov@imec.msu.ru}
\author{Alexander P. Seyranian}
\email{seyran@imec.msu.ru}
 \affiliation{Institute of Mechanics, Moscow State Lomonosov
 University\\ Michurinskii pr.~1, 119192 Moscow, Russia}%

\date{\today}

\begin{abstract}
A wave function picks up, in addition to the dynamic phase, the
geometric (Berry) phase when traversing adiabatically a closed
cycle in parameter space. We develop a general multidimensional
theory of the geometric phase for (double) cycles around
exceptional degeneracies in non-Hermitian Hamiltonians. We
show that the geometric phase is exactly $\pi$ for symmetric
complex Hamiltonians of arbitrary dimension and for nonsymmetric
non-Hermitian Hamiltonians of dimension 2. For nonsymmetric
non-Hermitian Hamiltonians of higher dimension, the geometric
phase tends to $\pi$ for small cycles and changes as the cycle
size and shape are varied. We find explicitly the leading
asymptotic term of this dependence, and describe it in terms of
interaction of different energy levels.
\end{abstract}


\maketitle

Non-Hermitian dissipative terms enter a quantum system Hamiltonian
when studying non-isolated systems, e.g., effective Hamiltonians
describing decay of unstable states. It turned out that the
non-Hermitian physics differs dramatically from the Hermitian
physics in the presence of degeneracies (energy level crossings),
even if the non-Hermitian system is close to the Hermitian
one~\cite{Berry2004, Heiss2004}. The most important degeneracy
intrinsic to non-Hermitian Hamiltonians is the exceptional point
(EP), at which two eigenvalues and corresponding eigenvectors
coalesce, as opposed to the diabolic point (DP) degeneracy of
Hermitian operators, at which the eigenvalues coalesce while the
eigenvectors remain different. EP degeneracies have been observed
in laser induced ionization of atoms~\cite{LatinneEtAl1995},
microwave cavities~\cite{BrentanoPhilipp1999,
DembowskiEtAl2001and2003}, in ``crystals of
light''~\cite{OberthalerEtAl1996}. Similar phenomena (where the
Hamiltonian is substituted by a different system operator) are
encountered in optics of absorptive media~\cite{BerryDennis2003},
acoustics~\cite{ShuvalovScott2000}, electronic
circuits~\cite{StehmannEtAl2004}, and mechanical
systems~\cite{SeyranianMailybaev2003, KirillovSeyranian2004}.

A wave function of a quantum system, whose parameters undergo
adiabatic cyclic evolution, acquires a complex factor dependent
only on the loop in parameter space and, thus, called geometric or
Berry phase~\cite{Berry1984}. Geometric phases in non-Hermitian
systems were studied in \cite{GarrisonWright1989,
Berry1990and1995, Massar1996, MondragonHernandez1996,
Heiss1999-2001, BerryDennis2003, KorschMossmann2003, KeckEtAl2003,
Heiss2004}. In such systems, it is important whether EP is inside
the closed path or not. For Hamiltonians given by specific
$2\times 2$ matrices, the geometric phase for a (double) cycle
around EP was found to be exactly $\pi$. Later this result was
verified experimentally in~\cite{DembowskiEtAl2001and2003}. So
far, EPs have been observed in decaying systems described by
symmetric effective Hamiltonians. This is the case when the
corresponding isolated system is time-reversal (described by a
real symmetric Hamiltonian). Time-irreversal interactions, e.g.,
with external magnetic field, break the symmetry of the effective
Hamiltonian.

We should note that the existing theoretical studies for the
geometric phase around EPs rely on the possibility of reducing the
system to the two-dimensional form. However, one should be aware
that the geometric phase is generally not preserved under such
reduction, as this reduction is given by a parameter dependent
change of basis. For example, we mention the change of geometric
phase under the parameter-dependent magnetic gauge
transformation~\cite{Berry1990and1995}.

In this paper, we develop a general multidimensional theory for
geometric phases around EPs. We show that, for symmetric complex
Hamiltonians of arbitrary dimension and for general non-Hermitian
Hamiltonians of dimension 2, the geometric phase is exactly $\pi$.
However, for nonsymmetric non-Hermitian Hamiltonians of higher
dimension, the geometric phase generally diverges from $\pi$ as
the cycle size increases. We find explicitly the leading term of
this divergence. It describes the background influence of energy
levels not involved in the EP degeneracy. We note that the
divergence from $\pi$ is related to irreversible Hermitian terms,
rather than to non-Hermitian dissipative terms.

\begin{figure}
\includegraphics[angle=0, width=0.27\textwidth]{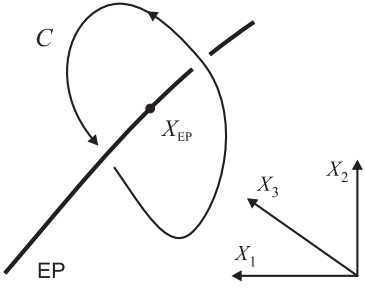}
\caption{Cycle around EP in parameter space.} \label{fig1}
\end{figure}

Let $H(X)$ be a non-Hermitian complex Hamiltonian smoothly
dependent on a vector of $m$ real parameters $X =
(X_1,\ldots,X_m)$. For simplicity, we consider Hamiltonians
represented by non-Hermitian complex matrices of arbitrary
dimension, but the results are valid in infinite dimensional case
as well. Let $E_n(X)$ be the eigenvalues of $H(X)$ (labeled $n$),
and $|\psi_n(X)\rangle$ be the corresponding eigenvectors. In
multiparameter space, a set of EPs defines a smooth surface of
codimension $2$~\cite{Arnold1983}. For clarity, we assume that the
number of parameters is three (then EPs form a curve), keeping in
mind that the results below are valid for any number of
parameters. Consider the EP curve, corresponding to the
coincidence of the levels $E_n = E_{n+1}$ and the eigenvectors
$|\psi_n(X)\rangle = |\psi_{n+1}(X)\rangle$. Let $C = \{X(t):\,0
\le t \le T\, \}$ be a cycle making one turn around this EP curve
in parameter space, see Fig.~\ref{fig1}. We assume that there are
no degeneracies (multiple eigenvalues) at points of the cycle $C$,
as well as there are no other EP curves inside $C$. We note that
EP is the only generic codimension 2 degeneracy for complex
non-Hermitian Hamiltonians smoothly dependent on
parameters~\cite{Arnold1983}. Thus, strictly speaking, EP is the
only degeneracy that can be encircled by $C$ in generic systems.

Let $|\Psi_n(0)\rangle = |\psi_n(X(0))\rangle$ and
$|\Psi_{n+1}(0)\rangle = |\psi_{n+1}(X(0))\rangle$ be the
interacting quantum states at $t = 0$. After traversing the cycle
$C$ once, the states interchange (up to the phase
multiplier)~\cite{Heiss1999-2001}. When making two turns, both
states return to their initial values picking up, in addition to
the usual dynamical phase $\delta_n = -\frac{1}{\hbar}\int_0^{2T}
E_n(t)dt$, a geometric phase $\gamma_n$~\cite{Berry1984}:
$|\Psi_{n,n+1}(2T)\rangle =
e^{i(\delta_n+\gamma_n)}|\Psi_{n,n+1}(0)\rangle$. Note that, due
to the interchanging of the states, we have $\delta_n =
\delta_{n+1}$ and $\gamma_n = \gamma_{n+1}$. For non-Hermitian
systems, the geometric phase is given by the
integral~\cite{GarrisonWright1989, Berry1990and1995}
    \begin{equation}
    \label{eq1.3}
    \gamma_n = \gamma_{n+1}
    = i\oint_{2C} \frac{\langle\widetilde{\psi}_n(X)|
    d\psi_n(X)\rangle}{\langle\widetilde{\psi}_n(X)|
    \psi_n(X)\rangle},
    \end{equation}
where $\langle\widetilde{\psi}_n(X)|$ is the left eigenvector
corresponding to $E_n(X)$. The integral in (\ref{eq1.3}) is
evaluated over the cycle $C$ traversed twice in the increasing
time direction (we denote this by $2C$). The right and left
eigenvectors are orthogonal at EP~\cite{Gantmacher1998}, which
means that the denominator of the integral expression in
(\ref{eq1.3}) is zero at EP.

First, consider complex symmetric Hamiltonians: $H(X) = H^T(X)$.
In this case the left and right eigenvectors are complex
conjugate: $\langle\widetilde{\psi}_{n,n+1}(X)| =
\overline{\langle\psi_{n,n+1}(X)|}$. By using this property, we
write (\ref{eq1.3}) in the form
    \begin{equation}
    \label{eq3.2}
    \begin{array}{rcl}
        \gamma_n & = & \displaystyle
        i\oint_{2C} \frac{\overline{\langle\psi_n(X)}|
        d\psi_n(X)\rangle}{\overline{\langle\psi_n(X)}|
        \psi_n(X)\rangle} \\[17pt]
        & = & \displaystyle
        \frac{i}{2}\oint_{2C}
        d\ln{\overline{\langle\psi_n(X)}|
        \psi_n(X)\rangle}.
    \end{array}
    \end{equation}
The phase $\gamma_n$ is equal to the change of the complex
quantity $\ln\overline{\langle\psi_n(X)}|\psi_n(X)\rangle$ over
the double cycle $2C$. This change depends only on the number of
turns made by the complex number
$\overline{\langle\psi_n(X)}|\psi_n(X)\rangle$ around zero in
complex plane, where zero is a branch point of the complex
logarithm function. In one turn around zero, the logarithm changes
by $\pm2\pi i$~\cite{Krantz1999}.

Since the geometric phase $\gamma_n$ does not depend on the form
and size of the cycle, we evaluate $\gamma_n$ by considering small
cycles $C$ around a point $X_\EP$ of the EP curve. At $X_\EP$, two
eigenvalues of the Hamiltonian $H_\EP = H(X_\EP)$ coalesce: $E_\EP
= E_n(X_\EP) = E_{n+1}(X_\EP)$. $E_\EP$ has a single eigenvector
$|\chi_0^\EP\rangle = |\psi_n(X_\EP)\rangle =
|\psi_{n+1}(X_\EP)\rangle$ and an associated vector
$|\chi_1^\EP\rangle$ determined by~\cite{SeyranianMailybaev2003}
    \begin{equation}
    \label{eq0.4}
    H_\EP
    |\chi_0^\EP\rangle
    {=} E_\EP
    |\chi_0^\EP\rangle,
    \ \,
    H_\EP
    |\chi_1^\EP\rangle
    {=} E_\EP
    |\chi_1^\EP\rangle
    {+}|\chi_0^\EP\rangle.
    \end{equation}
The left eigenvector $\langle\widetilde{\chi}_{0}^\EP| =
\langle\widetilde{\psi}_n(X_\EP)| =
\langle\widetilde{\psi}_{n+1}(X_\EP)|$ and associated vector
$\langle\widetilde{\chi}_{1}^\EP|$ are determined by
    \begin{equation}
    \label{eq0.10}
    \langle\widetilde{\chi}_0^\EP|
    H_\EP
    {=} E_\EP
    \langle\widetilde{\chi}_0^\EP|,
    \ \,
    \langle\widetilde{\chi}_1^\EP|
    H_\EP
    {=} E_\EP
    \langle\widetilde{\chi}_1^\EP|
    {+}\langle\widetilde{\chi}_0^\EP|.
    \end{equation}
Recall that $\langle\widetilde{\chi}_0^\EP |\chi_0^\EP\rangle =
0$. Additionally, we impose the normalization conditions
    \begin{equation}
    \label{eq0.10b}
    \langle\widetilde{\chi}_1^\EP
    |\chi_0^\EP\rangle =
    \langle\widetilde{\chi}_0^\EP
    |\chi_1^\EP\rangle = 1,\quad
    \langle\widetilde{\chi}_1^\EP
    |\chi_1^\EP\rangle = 0.
    \end{equation}

In the neighborhood of $X_\EP$, we have~\cite{SeyranianEtAl2005}
    \begin{equation}
    \label{eq3.2c}
    \begin{array}{l}
        |\psi_{n,n+1}(X)\rangle
        = |\chi_0^\EP\rangle
        {\pm}\sqrt{\mu}\,|\chi_1^\EP\rangle
        {+}o(\sqrt{\|X-X_\EP\|}),\\[5pt]
        \langle\widetilde{\psi}_{n,n+1}(X)|
        = \langle\widetilde{\chi}_{0}^\EP|
        {\pm}\sqrt{\mu}\,
        \langle\widetilde{\chi}_{1}^\EP|
        {+}o(\sqrt{\|X-X_\EP\|}),
    \end{array}
    \end{equation}
where $\mu$ is the linear scalar function of parameters
    \begin{equation}
    \label{eq3.2d}
    \mu(X) = \sum_{j = 1}^m
    {\langle\widetilde{\chi}_0^\EP}
    |\partial H/\partial X_j
    |\chi_0^\EP\rangle
    (X_j{-}X_j^\EP).
    \end{equation}
with the derivatives taken at $X_\EP$; the equation $\mu = 0$
gives the tangent of the EP curve in parameter
space~\cite{SeyranianEtAl2005}. By using (\ref{eq0.10b}),
(\ref{eq3.2c}), and the property
$\langle\widetilde{\chi}_{0,1}^\EP| =
\overline{\langle\chi_{0,1}^\EP|}$ for symmetric matrices, we
obtain
    \begin{equation}
    \label{eq3.2b}
    \overline{\langle\psi_n(X)}|\psi_n(X)\rangle =
    2\sqrt{\mu}+o(\sqrt{\|X-X_\EP\|}).
    \end{equation}
The complex number $\mu$ makes one turn around zero in complex
plane for one cycle $C$ in parameter space. Hence,
$\overline{\langle\psi_n(X)}|\psi_n(X)\rangle$ makes a single
closed loop around zero in complex plane for the double cycle
$2C$. As a result, the complex logarithm function in (\ref{eq3.2})
changes by $\pm2\pi i$, and we obtain $\gamma_n = \pm\pi$. The
sign depends on the direction of the cycle in complex plane; it
does not influence the final result, since the phase is determined
up to the additional term $2\pi k$ for any integer $k$.

There is the geometric phase analogy between EPs of complex
symmetric Hamiltonians and DPs of real symmetric Hamiltonians. For
real symmetric Hamiltonians, just like for complex symmetric
Hamiltonians, the geometric phase is ``produced'' only by the
degeneracies: it is $\pi$ if the degeneracy is encircled, and zero
otherwise~\cite{Berry1990and1995}. Such phases, which do not
depend on the shape (geometry) of the cycle, are called
topological~\cite{BohmEtAl2003}. The major difference between
complex and real cases is that the cycle should be traversed twice
for EP and once for DP. When a complex symmetric perturbation is
given to a real symmetric Hamiltonian, DP splits into two
EPs~\cite{KirillovEtAl2004}. One can say that each EP takes half
of the geometric phase of DP (counted per single cycle).

Now, consider nonsymmetric non-Hermitian Hamiltonians. We study
the local structure of the EP degeneracy by means of the versal
deformation theory of matrices~\cite{Arnold1983,
Mailybaev2000and2001}. The eigenvectors $|\psi_n(X)\rangle$ and
$|\psi_{n+1}(X)\rangle$ are nonsmooth functions of parameters at
$X_\EP$. However, together they define a two-dimensional invariant
linear subspace, which smoothly depends on parameters. This
invariant linear subspace can be given by two vectors
$|\chi_0(X)\rangle$ and $|\chi_1(X)\rangle$ smoothly dependent on
parameters: $|\chi_{0,1}(X)\rangle$ are linear combinations of
$|\psi_n(X)\rangle$ and $|\psi_{n+1}(X)\rangle$ and satisfy the
equations~\cite{Mailybaev2000and2001}
    \begin{equation}
    \label{eq0.1}
    \begin{array}{l}
        H(X)|\chi_0(X)\rangle
        = s(X)|\chi_0(X)\rangle+p(X)|\chi_1(X)\rangle,\\[5pt]
        H(X)|\chi_1(X)\rangle
        = s(X)|\chi_1(X)\rangle+|\chi_0(X)\rangle.
    \end{array}
    \end{equation}
Here $s(X) = (E_n(X)+E_{n+1}(X))/2$ and $p(X) =
(E_{n+1}(X)-E_{n}(X))^2/4$ are smooth scalar functions. At $X =
X_\EP$, where $s(X_\EP) = E_\EP$ and $p(X_\EP) = 0$, (\ref{eq0.1})
yield the Jordan chain equations (\ref{eq0.4}). Hence,
$|\chi_0(X_\EP)\rangle = |\chi_0^\EP\rangle$ is the eigenvector
and $|\chi_1(X_\EP)\rangle = |\chi_1^\EP\rangle$ is the associated
vector of the double eigenvalue $E_\EP$. By means of
(\ref{eq0.1}), the eigenvalues $E_{n,n+1}(X)$ and corresponding
eigenvectors are found as
    \begin{equation}
    \label{eq0.3}
    \begin{array}{rcl}
        E_{n,n+1}(X) & = & s(X)\pm\sqrt{p(X)},\\[3pt]
        |\psi_{n,n+1}(X)\rangle
        & = & |\chi_0(X)\rangle\pm\sqrt{p(X)}|\chi_1(X)\rangle,
    \end{array}
    \end{equation}
where two Riemann sheets of the complex square root correspond to
$E_n(X)$ and $E_{n+1}(X)$. We remark that the function $\mu(X)$ in
(\ref{eq3.2d}) is the linearization of $p(X)$ at $X_\EP$.
Similarly, the vectors $\langle\widetilde{\chi}_{0,1}(X)|$ are
introduced for the left eigenspace: they determine the left
eigenvectors as
    \begin{equation}
    \label{eq0.5}
    \langle\widetilde{\psi}_{n,n+1}(X)|
    = \langle\widetilde{\chi}_0(X)|
    \pm\sqrt{p(X)}\langle\widetilde{\chi}_0(X)|,
    \end{equation}
and satisfy the orthonormality conditions
    \begin{equation}
    \label{eq0.6}
    \begin{array}{c}
        \langle\widetilde{\chi}_0(X)|\chi_0(X)\rangle
        = \langle\widetilde{\chi}_1(X)|\chi_1(X)\rangle = 0,\\[3pt]
        \langle\widetilde{\chi}_1(X)|\chi_0(X)\rangle
        = \langle\widetilde{\chi}_0(X)|\chi_1(X)\rangle = 1.
    \end{array}
    \end{equation}
At EP, $\langle\widetilde{\chi}_{0}^\EP|
=
\langle\widetilde{\chi}_{0}(X_\EP)|$ is the left eigenvector and
$\langle\widetilde{\chi}_{1}^\EP|
=
\langle\widetilde{\chi}_{1}(X_\EP)|$ is the left associated
vector.

By using (\ref{eq0.3})--(\ref{eq0.6}) in (\ref{eq1.3}), we obtain
    \begin{equation}
    \label{eq2.10}
    \begin{array}{l}
        \displaystyle
        \gamma_n =
        \frac{i}{2}\oint_{2C}d\ln\sqrt{p(X)}\\[17pt]
        \displaystyle
        \ \,+\,i\oint_{2C}
        \frac{\langle\widetilde{\chi}_0(X)|d\chi_0(X)\rangle
        {+}p(X)\langle\widetilde{\chi}_1(X)|d\chi_1(X)\rangle}{2\sqrt{p(X)}}
        \\[17pt]
        \displaystyle
        \ \,+\,\frac{i}{2}\oint_{2C}
        \big(\langle\widetilde{\chi}_0(X)|d\chi_1(X)\rangle
        {+}\langle\widetilde{\chi}_1(X)|d\chi_0(X)\rangle\big).
    \end{array}
    \end{equation}
The double cycle $2C$ corresponds to a single cycle of the square
root $\sqrt{p(X)}$ around zero in complex plane. Hence, the first
integral in (\ref{eq2.10}) equals $\pm 2\pi i$, where the sign
depends on the direction of the cycle in complex plane. The second
integral in (\ref{eq2.10}) vanishes, since the square root in the
denominator has opposite signs when traversing the first and
second cycles. Finally, the third integral is the same for the
first and second cycles. As a result, we have
    \begin{equation}
    \label{eq0.7}
    \gamma_n
    = \pm\pi+i\oint_{C}
    \big(\langle\widetilde{\chi}_0(X)|d\chi_1(X)\rangle
    {+}\langle\widetilde{\chi}_1(X)|d\chi_0(X)\rangle\big).
    \end{equation}
Remark that the integral in (\ref{eq0.7}) is taken over one cycle
$C$ in the increasing time direction.

First, consider Hamiltonians given by $2\times2$ general complex
matrices. According to (\ref{eq0.6}), the $2\times2$ matrix
$|1\rangle\langle\widetilde{\chi}_1(X)|
+|2\rangle\langle\widetilde{\chi}_0(X)|$ is the inverse of
$|\chi_0(X)\rangle\langle1|+|\chi_1(X)\rangle\langle2|$, where
$|1\rangle = (1,\,0)$ and $|2\rangle = (0,\,1)$ are the unit
vectors. Hence, components of the vectors
$\langle\widetilde{\chi}_{0,1}(X)|$ can be expressed explicitly in
terms of the components of $|\chi_{0,1}(X)\rangle$. By using these
expressions, we transform the integral in (\ref{eq0.7}) to the
form $\oint_{C}d\ln\det(|\chi_0(X)\rangle\langle1|
+|\chi_1(X)\rangle\langle2|)$; it vanishes since the $2\times2$
matrix $|\chi_0(X)\rangle\langle1| +|\chi_1(X)\rangle\langle2|$ is
everywhere nonsingular by definition. Hence, for $2\times2$
general non-Hermitian Hamiltonians, the geometric phase equals
$\pm\pi$ and does not depend on the loop shape, similar to the
case of symmetric complex Hamiltonians. This result
justifies the existence of topological indices describing the
polarization ellipses around C points in crystal
optics~\cite{BerryDennis2003}.

For multidimensional non-Hermitian Hamiltonians, the integral in
(\ref{eq0.7}) is generally nonzero. Consider a cycle $C = \{X(t) =
X_\EP+\varepsilon\widehat{X}(t):\,0 \le t \le T\, \}$ making one
turn around EP, where $\varepsilon$ is a small positive parameter
controlling size of the cycle. Formulae for derivatives of
$|\chi_{0,1}(X)\rangle$ and $\langle\widetilde{\chi}_{0,1}(X)|$ at
$X_\EP$ are provided by the versal deformation
method~\cite{Mailybaev2000and2001}. By using these formulae in
(\ref{eq0.7}), we obtain the asymptotic expression
    \begin{equation}
    \label{eq0.8}
    \gamma_n
    = \pm\pi
    +ia\varepsilon^2+O(\varepsilon^3),
    \end{equation}
where the complex constant $a$ is given by the integral
    \begin{equation}
    \label{eq0.9}
    \begin{array}{l}
        \displaystyle
        \!\!\!\!a = \oint_{C}
        \Big(2\langle\widetilde{\chi}_0^\EP|
        H_1(G^{-3}{-}|\chi_1^\EP\rangle
        \langle\widetilde{\chi}_1^\EP|)
        dH_1|\chi_0^\EP\rangle\\[10pt]
        \!\!+\,\langle\widetilde{\chi}_0^\EP|
        H_1G^{-2}dH_1
        |\chi_1^\EP\rangle
        {+}\langle\widetilde{\chi}_1^\EP|
        H_1G^{-2}dH_1
        |\chi_0^\EP\rangle
        \Big).
    \end{array}
    \end{equation}
Here $H_1(\widehat{X}) = \sum_{j = 1}^m(\partial H/\partial
X_j)\widehat{X}_j$ and $dH_1(\widehat{X}) = \sum_{j =
1}^m(\partial H/\partial X_j)d\widehat{X}_j$ with the partial
derivatives taken at $X_\EP$, and $G = H_\EP-E_\EP I
+|\chi_1^\EP\rangle \langle\widetilde{\chi}_1^\EP|$ is a
nonsingular matrix ($I$ is the identity operator). The correction
term $ia\varepsilon^2$ is determined by the information about the
system at EP (this includes eigenvectors, associated vectors, and
first derivatives of the Hamiltonian with respect to parameters)
and by the cycle shape $\widehat{X}(t)$.
Details of the derivation of (\ref{eq0.9}) will appear
elsewhere~\cite{MailybaevKirillov2005}.

The physical meaning of the constant (\ref{eq0.9}) can be
understood by using the eigenvector expansion of the unity and of
the Hamiltonian at EP:
    \begin{equation}
    \label{eq0.11b}
    I
    = |\chi_0^\EP\rangle
    \langle\widetilde{\chi}_1^\EP|
    +|\chi_1^\EP\rangle
    \langle\widetilde{\chi}_0^\EP|
    +\sum_{k \ne n,n+1}
    |\psi_k^\EP\rangle
    \langle\widetilde{\psi}_k^\EP|,
    \end{equation}
    \begin{equation}
    \label{eq0.11}
    \begin{array}{l}
        H_\EP
        = |\chi_0^\EP\rangle
        \langle\widetilde{\chi}_0^\EP|
        {+}E_\EP\big(
        |\chi_0^\EP\rangle
        \langle\widetilde{\chi}_1^\EP|
        {+}|\chi_1^\EP\rangle
        \langle\widetilde{\chi}_0^\EP|
        \big)
        \\[10pt]\displaystyle
        \qquad\ \ \ +\sum_{k \ne n,n+1}
        E_k^\EP
        |\psi_k^\EP\rangle
        \langle\widetilde{\psi}_k^\EP|,
    \end{array}
    \end{equation}
where $E_k^\EP = E_k(X_\EP)$, $|\psi_k^\EP\rangle =
|\psi_k(X_\EP)\rangle$, and $\langle\widetilde{\psi}_k^\EP| =
\langle\widetilde{\psi}_k(X_\EP)|$. Here we assume the
normalization condition for the left and right eigenvectors
$\langle\widetilde{\psi}_{k}^\EP |\psi_k^\EP\rangle = 1$. Recall
that $\langle\widetilde{\psi}_{k}^\EP |\chi_{0,1}^\EP\rangle =
\langle\widetilde{\chi}_{0,1}^\EP |\psi_k^\EP\rangle = 0$ and
$\langle\widetilde{\psi}_{k'}^\EP |\psi_k^\EP\rangle = 0$ if $k
\ne k'$. Expression (\ref{eq0.11}) represents the transformation
of $H_\EP$ to the canonical Jordan form written in terms of
eigenvectors and associated vectors~\cite{Gantmacher1998}. By
substituting (\ref{eq0.11b}) and (\ref{eq0.11}) into the
expression for the matrix $G$, after a series of manipulations, we
transform (\ref{eq0.9}) to
    \begin{equation}
    \label{eq0.12}
    \begin{array}{l}
        \displaystyle
        a = \sum_{k \ne n,n+1}\oint_{C}\left(
        2\,\frac{\langle
        \widetilde{\chi}_0^\EP|
        H_1|\psi_k^\EP\rangle
        \langle
        \widetilde{\psi}_k^\EP|dH_1
        |\chi_0^\EP\rangle}{
        (E_k^\EP
        -E_\EP)^3}\right.\\[15pt]
        \displaystyle
        \qquad\qquad\qquad
        +\,\frac{\langle
        \widetilde{\chi}_1^\EP|
        H_1|\psi_k^\EP\rangle
        \langle
        \widetilde{\psi}_k^\EP|dH_1
        |\chi_0^\EP\rangle}{
        (E_k^\EP
        -E_\EP)^2}\\[12pt]
        \displaystyle
        \qquad\qquad\qquad
        \left.+\,\frac{\langle
        \widetilde{\chi}_0^\EP|
        H_1|\psi_k^\EP\rangle
        \langle
        \widetilde{\psi}_k^\EP|dH_1
        |\chi_1^\EP\rangle}{
        (E_k^\EP
        -E_\EP)^2}\right).
    \end{array}
    \end{equation}

The terms $\langle\widetilde{\chi}_{0,1}^\EP|
H_1|\psi_k^\EP\rangle$ and $\langle \widetilde{\psi}_k^\EP|dH_1
|\chi_{0,1}^\EP\rangle$ describe the interaction of the degenerate
level $E_\EP$ with the levels $E_k$, $k \ne n,n+1$ at the EP.
Thus, the change of the geometric phase with the cycle size and
shape variation is due to the influence of the energy levels not
involved in the EP degeneracy. One can see that, if the difference
$E_k^\EP -E_\EP$ is big, the influence of the level $E_k$ is
proportional to $(E_k^\EP -E_\EP)^{-2}$ and can be neglected.
However, if $E_k^\EP -E_\EP$ is small, the change of the geometric
phase due to the interaction with $E_k$ grows proportionally to
$(E_k^\EP -E_\EP)^{-3}$ and may be big. In the extreme case
$E_k^\EP -E_\EP \rightarrow 0$, i.e., near the triple degeneracy
$E_n = E_{n+1} = E_k$, we have $a \rightarrow \infty$. Hence,
triple degeneracies require special investigation.

Thus, for nonsymmetric non-Hermitian Hamiltonians, the deviation
of the geometric phase from $\pi$ is the multidimensional
phenomenon, which cannot be captured in two-dimensional
approximations. Asymptotic expression (\ref{eq0.8}) with the
coefficient (\ref{eq0.9}) for the correction term was confirmed by
numerical simulations for particular Hamiltonians of dimensions 3
and 4. We believe that this change of the geometric phase, which
is intrinsic to nonsymmetric non-Hermitian Hamiltonians, can be
verified in future experiments. For example, this effect should
exist when studying the decay of nearly degenerate unstable states
for time-irreversal systems (Hamiltonians must have nonsymmetric
Hermitian terms). Probably, the experimental approach
of~\cite{DembowskiEtAl2001and2003} can be used for this purpose if
one manages to break the symmetry of the effective Hamiltonian in
a controllable way.

Based on the expansions of eigenvectors near EP we have shown that
in the general case the geometric phase integral can be evaluated
by methods of complex analysis. This can be regarded as a response
to Arnold~\cite{Arnold1995} who suggested to develop a theory of
"residues" to calculate the Berry phase.

This work has been supported by the research grants RFBR
03-01-00161, CRDF-BRHE Y1-MP-06-19, and CRDF-BRHE Y1-M-06-03.

\end{document}